\documentclass[twocolumn,showpacs,
amsmath,amssymb]{revtex4}
\usepackage{amssymb}
\usepackage[dvips]{graphicx}
\begin{document}

\title{Unconventional proximity effects and pairing symmetries in cuprates caused by
                        conventional phonons}
\author{A. S. Alexandrov}

\affiliation{Department of Physics, Loughborough University,
Loughborough LE11 3TU, United Kingdom\\}

\begin{abstract}
Giant and nil proximity effects and unconventional symmetry of
cuprate superconductors are explained as a result of the strong
Fr\"ohlich  interaction of holes with  c-axis polarised optical
phonons acting together with an anisotropic nonlocal deformation
potential.

\end{abstract}


\maketitle

A  number of  observations point to the possibility that many
high-$T_c$ cuprate superconductors may not be conventional
Bardeen-Cooper-Schrieffer (BCS) superconductors, but rather derive
from the Bose-Einstein condensate (BEC) of real-space pairs, which
are mobile small bipolarons \cite{alerev}. A possible fundamental
origin of such strong departure of the cuprates from conventional
BCS behaviour is the Fr\"ohlich electron-phonon interaction (EPI) of
the order of 1 eV \cite{ale96}, routinely neglected in the Hubbard U
and t-J models of cuprate superconductors \cite{ran}. This
interaction with c-axis polarized optical phonons is virtually
unscreened since the upper limit for the out-of-plane plasmon
frequency in cuprates is well below the characteristic frequency of
optical phonons.  Due to a poor screening, the magnetic interaction
remains small compared with the Fr\"ohlich EPI at any doping of
cuprates. Combined with on-site repulsive correlations (Hubbard U)
and short-range deformation (i.e acoustic phonon) and Jahn-Teller
\cite{mul} EPIs, the unscreened long-range Fr\"ohlich EPI     binds
oxygen holes into superlight intersite bipolarons \cite{ale96,kor},
which bose-condense at very high temperatures \cite{hag}.

Experimental evidence for  exceptionally strong EPI is now
overwhelming \cite{alerev}, so that the bipolaronic superfluid
charged Bose gas could be a feasible alternative to the BCS-like
scenarios of cuprates. The bipolaron theory predicted such key
features of cuprate superconductors as anomalous upper critical
fields \cite{alehc2}, spin and charge pseudogaps \cite{gap}, and
unusual isotope effects \cite{iso} later discovered experimentally.
The theory accounts for normal state kinetics, including the
anomalous Hall-Lorenz numbers, high $T_c$ values, specific heat
anomalies of cuprates (for a review see \cite{alerev}), and also for
the d-wave checkerboard order parameter \cite{wave}, the normal
state Nernst effect \cite{nernst}  and diamagnetism \cite{dia}.

Here I argue that the Bose-condensate tunneling into a normal
cuprate semiconductor \cite{prox} accounts for both nil \cite{nil}
and giant \cite{gia} proximity effects discovered experimentally in
cuprates. Several groups reported that in the cuprate SNS junctions
supercurrent can run through normal N-barriers as thick as 10 nm, if
the barriers are made from a non-superconducting cuprate layer.
Since the c-axis coherence length is only about 0.1 nm, these
observations are in a strong conflict with the standard BCS picture.
Using the advanced molecular beam epitaxy, Bozovic et al. \cite{gia}
demonstrated that the giant proximity effect (GPE) is intrinsic,
rather than  caused by any inhomogeneity of the barrier such as
stripes, superconducting "islands", etc.

This unusual effect can be broadly understood as the bipolaron BEC
tunnelling into a cuprate semiconductor, Fig.1 \cite{prox}. The
condensate wave function, $\psi(z)$, is described by the
Gross-Pitaevskii equation \cite{pit} as
\begin{equation}
{\hbar^2\over{2m^{**}}} {d^2 \psi(z)\over{dz^2}}= [V|\psi(z)|^2-\mu]
\psi(z),
\end{equation}
where $V$ is a short-range repulsion of bosons, and $m^{**}$ is the
boson mass along the direction of tunnelling $z$. If the normal
barrier, at $z > 0$, is an underdoped cuprate semiconductor above
its transition temperature,  the chemical potential $\mu$ is found
below the quasi-2D bosonic band at a very small energy, Fig.1,
\begin{equation}
\epsilon= - k_B T \ln[1- \exp(-T_0/T)].
\end{equation}
Here $T_0= \pi \hbar^2 x^\prime /k_B m^{**} >> T^\prime_c$  is much
higher than the transition temperature, $T^\prime_c$ , of the
barrier with the carrier density $x^\prime$. Hence Eqs.(1,2) predict
the occurrence of a new length scale,
$\hbar/(2m^{**}\epsilon)^{1/2}$, which is much larger than the
zero-temperature coherence length in a wide temperature range
$T^\prime_c < T < T_0$, accounting for GPE.

\begin{figure}
\begin{center}
\includegraphics[angle=-90,width=0.55\textwidth]{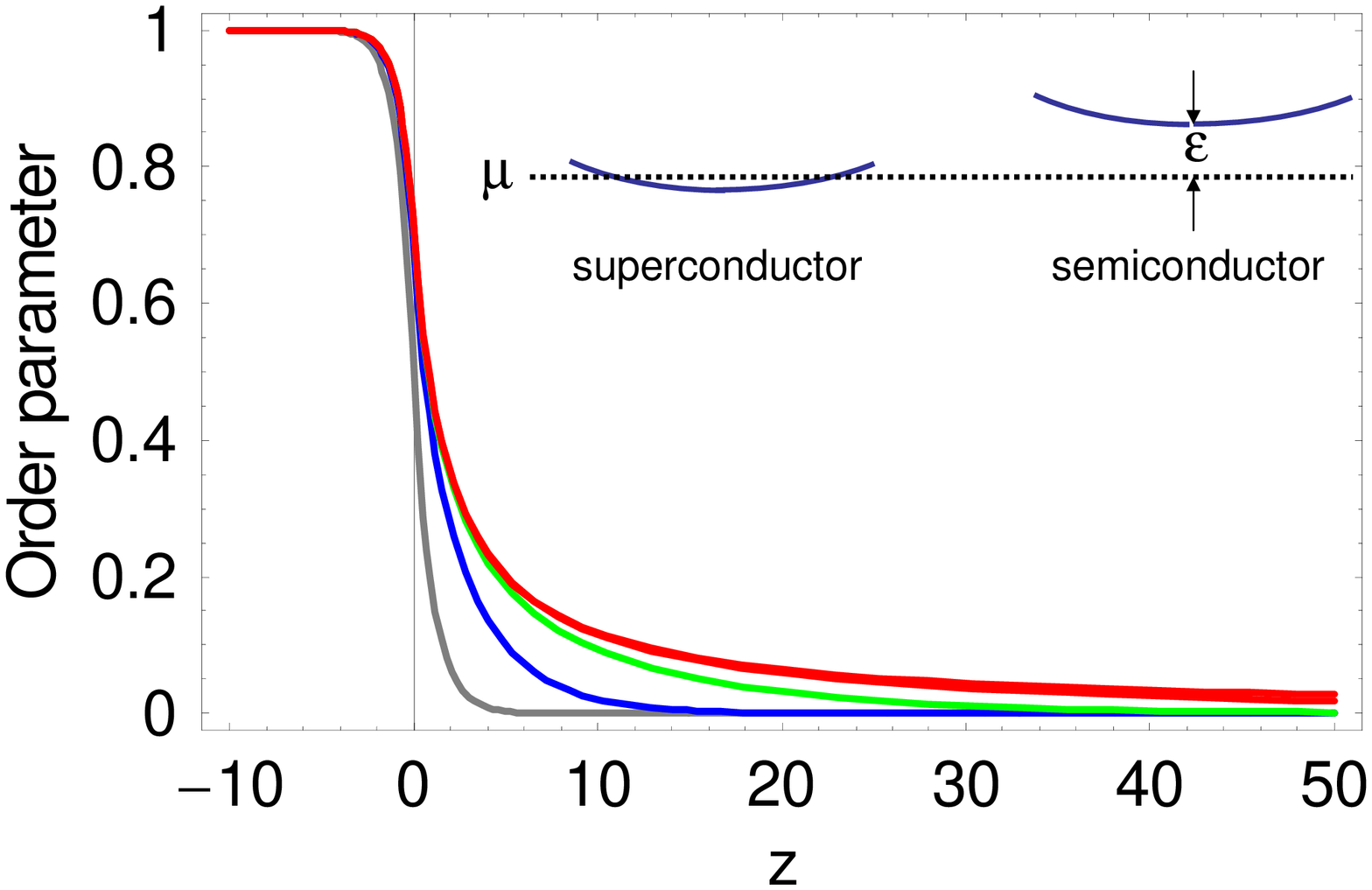}
\vskip -0.5mm \caption{BEC order-parameter near the SN boundary for
different doping levels of the normal barrier at z $> 0$ (z is
measured in units of the zero-temperature coherence length of the
superconductor, the upper curve corresponds to a smallest
$\epsilon$, after Ref. \cite{prox}.  The chemical potential is found
above the boson band-edge due to the boson-boson repulsion in
cuprate superconductors and below the edge in   cuprate
semiconductors with low doping.}
\end{center}
\end{figure}

The physical reason why the quasi-2D bosons display a large
normal-state coherence length, whereas 3D Bose-systems (or any-D
Fermi-systems) at the same values of parameters do not, originates
in the large density of states (DOS) near the band edge of
two-dimensional bosons compared with 3D DOS. Since DOS is large, the
chemical potential is pinned near the edge with the exponentially
small  magnitude, $\epsilon$ , at $T <T_0$. Importantly if the
barrier is undoped ($x^\prime\rightarrow 0$)  $\epsilon$  becomes
large at any finite temperature, which explains the nil proximity
effect \cite{nil} in the case of the undoped insulating barrier.

 I also argue that a huge  anisotropy of  the sound speed
in layered superconductors makes the acoustic phonon-mediated
attraction of electrons non-local in space providing unconventional
Cooper pairs with a nonzero orbital momentum in the weak-coupling
BCS regime \cite{wave2}. As a result of this anisotropy quasi-two
dimensional charge carriers weakly coupled with acoustic phonons
undergo a quantum phase transition from a conventional s-wave to an
unconventional d-wave superconducting state with less carriers per
unit cell. In the opposite strong-coupling regime rotational
symmetry breaking appears as a result of a reduced Coulomb repulsion
between unconventional bipolarons compared with the repulsion of
less extended s-wave pairs dismissing thereby some constraints on
unconventional (internal) symmetry of preformed pairs in the BEC
limit as well.

 The conventional acoustic phonons, and not superexchange \cite{ran},
 are shown to be responsible for the unconventional symmetry of cuprate superconductors, where the on-site Coulomb repulsion is large \cite{wave2}.

  I thank  A. F. Andreev, J. P. Hague,   V. V. Kabanov, P. E. Kornilovitch, and J. H. Samson for valuable discussions, and acknowledge support by EPSRC (UK) (grant
nos. EP/C518365/1, EP/D035589/1, and EP/D07777X/1).

\end{document}